# Development of a radio-detection array for the observation of UHE neutrino induced showers

Daniel Ardouin[1], Didier Charrier[1], Pascal Lautridou[1], Olivier Martineau-Huynh[2], Olivier Ravel[1], Xiang-Ping Wu[3], Meng Zhao[3]

(1) SUBATECH, IN2P3-CNRS- University of Nantes, E.Mines, Nantes, France

(2) LPNHE, IN2P3-CNRS, University Paris VI, France

(3) NAOC, Beijing, China

Abstract:
The recent demonstration by the CODALEMA Collaboration of the ability of the radio-detection technique for the characterization of ultra-high energy cosmic-rays (UHECR) calls for the use of this powerful method for the observation of UHE neutrinos. For this purpose, an adaptation of the already existing 21CM - Array in China, is presently under achievement. In an exceptionally low electromagnetic noise level, 10160 log-periodic 50-200 MHz antennas sit along two high altitude valleys, surrounded by mountain chains. This lay-out results in 30-60 km effective rock thicknesses for ν interactions with low incidence trajectories along the direction of two 4-6 km baselines. We will present first in-situ radio measurements demonstrating that this environment shows particularly favourable physical conditions for the observation of electromagnetic decay signals of τ's leptons originating from the interaction of $10^{17-20}$ eV $\nu_\tau$ neutrinos.

**1. Objectives of the neutrino experiment**

High energy astrophysical neutrinos of hadronic origin are believed to result from interaction of accelerated particles with radiation or gas. These scenarios are expected to happen in environments like jets of AGN, GRB sources or those resulting from the scattering of cosmic rays with background radiation. Among particles and gamma-rays resulting from the decay of produced pions, only neutrinos appear as possible direct messengers for high energy sources of radiation beyond several Mpc, due to the interacting contribution of the microwave background radiation. Their observation would obviously bring an interesting opening in the context of the recent observation of a cut-off in the spectrum above $10^{20}$ eV.

On the experimental point of view, the expected event rates of the order of less than 1/year/km³ at UHE, imply the use of huge detection volumes. Though several operating or designed detectors with ~ km³ sensitive volume, using optical techniques, already exist, none of them has shown evidence for a ν of astrophysical origin yet, and only upper limits in the range ~ $10^{-7}$ to $10^{-6}$ GeV.cm⁻².s⁻¹.sr⁻¹ have been set by the late experiments, very useful for the existing models.

Hence, the use of high density target matter, like rocks, has been proposed by several authors [1], and seems better adapted in bringing neutrino mean-free paths of the order of several hundreds of km.

A first constraint to overcome an observable energy of $\sim 10^{15}$ eV is the choice of a restricted neutrino journey inside the Earth acting as a *neutrino converter*. A second constraint is a large enough surrounding *detection space* in order to observe the (lepton) decay-products. Both constraints can be satisfied by observing lepton-decay air showers developing in a valley area surrounded (or obscured) by high mountain chains used as neutrino interaction medium. In such a case, only the escaping $\tau$ lepton, resulting from a neutrino-nucleon interaction, possesses a disintegration length ($\Gamma \sim 50$ m x E[PeV]) compatible with a usable detection space.

One should add that observing showers from a source behind a mountain has also the advantage of minoring contamination from "direct" cosmic rays showers by means of an elevation analysis.

This contribution will present the choice of the 21CMA set-up as an interesting site combining these conditions while a radio technique method is proposed for the characterization of the $\tau$ decay-showers.

## 2. Specific advantages of the radio-detection method for a neutrino detection experiment

In association with the above neutrino target specifications, large effective *detection* volumes ( > 100 km$^3$) can be met by using the pioneering radio technique initially proposed by Askar'yan [2]. Validated recently with accelerator experiments on dense materials, this challenging technique is used by several experiments [3-7] observing the radio Cerenkov radiation following neutrino interaction in ice or salt. It is also more effective to fully exploit the predicted coherence effect in this wavelength domain.

An additional specific advantage of the radio-technique is its duty cycle close to 100 %.

The demonstrated sensitivity of the radio detection method up to large distances from the emitting sources bears two important consequences : (*i*) an improved detection acceptance which can partially fit the importance of long decay-lengths expected for high energy $\tau$'s ; ( *ii*) the sampling of a longer phase of the shower development.

The CODALEMA collaboration has firmly demonstrated the ability of a radio array to measure the *same* air- shower event as a ground detector array using an appropriate signal processing [8]. At the same time, this collaboration has shown that signal recognition algorithms have advanced enough to be confident in self- triggered radio arrays [9, 10]

## 3. The unique performances of the 21 CMA as a neutrino telesope

Our experimental site is the 21CM Array [11], located in the mountainous Xinjiang area of northern China, recently set-up for the study of the 21 cm emission from neutral hydrogen at the re-ionization epoch of the Universe.

The 21CM Array consists of 10287 [50-200 MHz] log-periodic antennas distributed over 80 pods of 127 antennas along 2 arms ( ~3 & 4 km long) oriented along the EW and NS directions respectively (Fig.1). The elevation directivity of the antennas is set close to the horizon line together with a radiation pattern lobe of +/- 45° azimuthally opening (at -3dB).

The electronics of each pod are powered by solar panels. After a 50-100 MHz band filter and a 60 dB amplifier, the signals are sent through optical transmitters into optical fibers to the control room where they are digitized by synchronized fast 8-bits 200 MS/s ADC's . For each pod, buffers are stored and on-line analyzed by a quad-processor PC.

This site presents two specific advantages which make up exceptional conditions for our purpose. The first one deals with the environment requirements (radiator and detection volumes) above presented. Situated at 2600-2700 m altitude, the site is indeed surrounded by up to 4200 m altitude mountain chains equivalent to a neutrino radiator converter volume of more than 20 000 km$^3$ /arm. For instance, the existence of strong natural depressions (the Tarim and Gobi deserts) in the SE to SW sector results in a 2500-3000 m high and 30-90 km thick "rock wall" representing a year - exposure of ~$2.10^{20}$ cm$^2$. s for a point-like source.

The second one, and probably world-unique, advantage, is the excellent noiseless radio environment offered by this remote site which shows (Fig.2) a quasi-absence of radio transmitters above 15 Mhz; thus, only the galactic (and atmospheric) noise will set the limits of observation.

*3.1. Operating procedure and first tests of UHECR radio detection at the 21 CMA*

We will operate in a self triggering mode as justified in [8-10]. The expected high multiplicity for showers propagating along (or close to) the direction of the two arm-valleys will favour this mode of operation while other (transverse) trajectories will be shielded by the environment. The trigger level is on-line adjusted by software with a duty cycle of 100%.

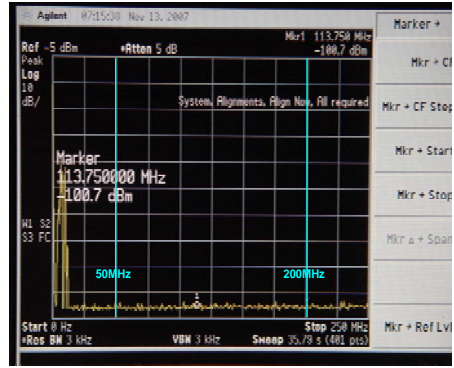

Fig.2 Typical sample of the radio environment signals at 21CMA observed on a spectrum analyzer. Note the absence of radio transmitters signal above 15 MHz (abscissa).

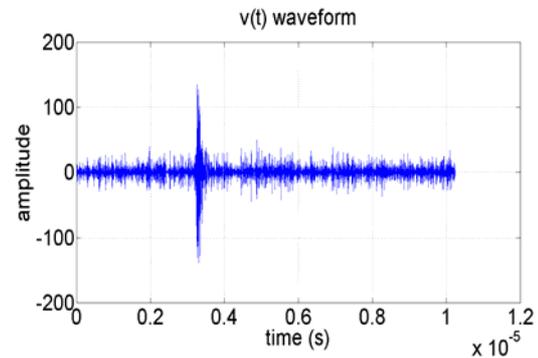

Fig.3. Example of observation of a transient signal at the 21 CMA resulting in a 3 antenna coincident event with a 50-100 MHz signal filtering. The voltage amplitude in ordinate is plotted in arbitrary units versus the time window in 10 μs units.

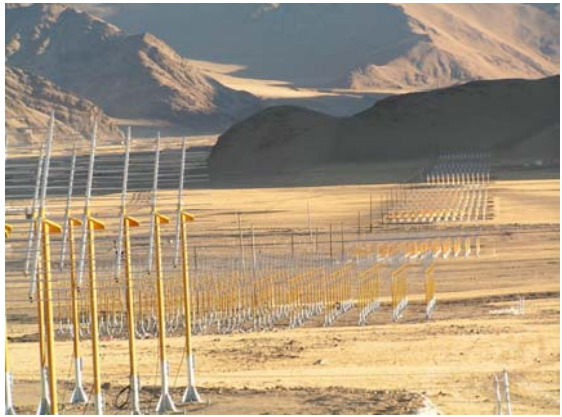

Fig.1. General view of the North-South arm of the 21 CMA.

From the average registered signal $v_{av}$ and its standard deviation σ, the trigger is set [8, 12] after a scan of each stored buffer: when a sample of events satisfies the condition $v > v_{av} + n\ \sigma$, (with n chosen according to the transient rate level), we perform a full waveform FFT analysis.

First test experiments (Fig. 3) have been recently carried out with a limited set of antennas belonging to different pods which clearly demonstrate the ability of the whole set-up to detect and correlate transient signals caused by "ordinary" cosmic rays.

*3.2. Criteria for τ showers discrimination*

To identify the air-shower trajectory and profile, the complete analysis of the time and amplitude distributions will be used in a customary way [9, 12]. In order to discriminate the ν–τ showers, we expect a correlation between the native decay-shower and the spreading of the signal and antenna multiplicity along the antenna arm. Simulations of such τ showers are in progress at different angles to estimate the amplitude of this effect.

Knowing the locations of the antenna along the slope of the arm and the time resolution of the radio signal ( ~ 5 ns) , the time sequence of the transient signals observed over successive fired antennas of one arm will also allow a precise ($\Delta\theta < 0.1°$) determination of the trajectory elevation. This will be used as a discrimination criterion against downwards usual CR showers using the known environment relief profile.

The available atmosphere range (5.7-10.5 km) from the mountain base (with the above average geometry characteristics) to the furthermost antennas allows the observation of the τ decay with an average counting efficiency of 1.2 to $6.10^{-3}$ between $10^{18}$ and $10^{20}$ eV.

## 4. Perspectives

From the above presented neutrino target characteristics and estimated detection efficiency, preliminary calculations indicate an averaged rate of ~ 1 event. $yr^{-1}$ for $10^{18}$ eV incident neutrinos assuming an energy flux $E^2\Phi = 4.10^{-8}$ [$GeV.s^{-1}.cm^{-2}.sr^{-1}$] as a benchmark. Simulations are currently being performed including the complete geometry of the detection area.
The most important conclusion of our study is the demonstration of the ability of the 21CMA whole-setup to characterise CR transient events.

The proposed neutrino conversion method associated with a radio-detection of resulting τ decay showers is capable of achieving a sensitivity similar that of existing (or planned) big detectors. The profile and elevation dependences of the transient radio-events observed along each antenna–arm are expected to reveal the $\nu_\tau$ origin (and energy dependence) of the inclined showers. The azimuthal exploration may also reveal the (non)-isotropic distribution of UHECR arrival directions.

The authors acknowledge the support of the France-China Particle Physics Laboratory (FCPPL).